\documentclass[11pt]{article}
\usepackage{graphicx}
\usepackage{latexsym}
\usepackage{epsfig,amsmath,amssymb,cite} 

\begin{document}

\newcommand{\field}[1]{\mathbb{#1}}
\newcommand{\BR}{\field{R}}
\newcommand\ZZ{\mathcal{Z}}
\newcommand{\arcsinh}{{\rm arcsinh}\,}
\newcommand{\arccosh}{{\rm arccosh}\,}
\newcommand{\eq}[2]{\begin{equation} #1 \label{#2} \end{equation}}
\newcommand\defeq{\mathrel{\mathop:}=}
\newcommand\nts{\!}
\newcommand\bns{\nts \nts \nts}

\newcommand{\eg}{{\it e.g.}}
\newcommand{\etal}{{\it et. al.}}
\newcommand{\ie}{{\it i.e.}}
\newcommand{\be}{\begin{equation}}
\newcommand{\dd}{\displaystyle}
\newcommand{\ee}{\end{equation}}
\newcommand{\bea}{\begin{eqnarray}}
\newcommand{\eea}{\end{eqnarray}}
\newcommand{\bef}{\begin{figure}}
\newcommand{\eef}{\end{figure}}
\newcommand{\bce}{\begin{center}}
\newcommand{\ece}{\end{center}}
\def\lsim{\mathrel{\rlap{\lower4pt\hbox{\hskip1pt$\sim$}}
    \raise1pt\hbox{$<$}}}         
\def\gsim{\mathrel{\rlap{\lower4pt\hbox{\hskip1pt$\sim$}}
    \raise1pt\hbox{$>$}}}         

\title{The Exact String Black-Hole \\ behind the hadronic Rindler horizon?}

\author{P.~Castorina$^{(a)}$\thanks{E-mail: paolo.castorina@ct.infn.it}, D.~Grumiller$^{(b)}$\thanks{E-mail: grumil@lns.mit.edu}, A.~Iorio$^{(c)}$\thanks{E-mail: iorio@ipnp.troja.mff.cuni.cz}}

\maketitle

\noindent $^{(a)}$ Dipartimento di Fisica, Universit\`a di Catania and INFN-Catania, \\
Via Santa Sofia 64, 95100 Catania - Italy\\
$^{(b)} $ Center for Theoretical Physics, Massachusetts Institute of Technology, \\
77 Massachusetts Ave., Cambridge, MA 02139 - USA \\
$^{(c)} $ Institute of Particle and Nuclear Physics, Charles University of Prague,  \\
V Hole\v{s}ovickach 2, 182 00 Prague 8 - Czech Republic

\begin{abstract}
The recently suggested interpretation \cite{Castorina:2007eb} of the universal hadronic freeze-out temperature
$T_f$ ($\simeq 170$ Mev) -- found for all high energy scattering processes that produce hadrons:
$e^+ e^-$, $p p$, $p \bar{p}$, $\pi p$, etc. and $NN^\prime$ (heavy-ion collisions) -- as a Unruh temperature
triggers here the search for the gravitational black-hole that in its near-horizon approximation better
{\it simulates} this hadronic phenomenon.
To identify such a black-hole we begin our gravity-gauge theory phenomenologies matching by asking the
question: which black-hole behind that Rindler horizon could reproduce the experimental behavior of $T_f (\sqrt{s})$
in $NN^\prime$, where $\sqrt{s}$ is the collision energy. Provided certain natural
assumptions hold, we show that the exact string black-hole turns out to be the best candidate (as it fits the available
data on $T_f (\sqrt{s})$) and that its limiting case, the Witten black-hole, is the unique candidate to explain
the constant $T_f$ for all elementary scattering processes at large energy. 
We also are able to propose an effective
description of the screening of the hadronic string tension $\sigma(\mu_b)$ due to the baryon density effects on $T_f$.
\end{abstract}

\section{Introduction}\label{se:1}

RHIC experimental data strongly suggest that, above the critical
temperature $T_c$ -- and up to $2.5 T_c$ -- QCD is a strongly
interacting system of quarks and gluons (for a brief review see, e.g.,
\cite{Kolb:2003dz}) -- a picture confirmed by lattice simulations \cite{Karsch:2000ps, Karsch:2003jg}.
Hence standard perturbative techniques fail to describe such a system -- similar
to a liquid with small shear viscosity $\eta$ \cite{Teaney:2003kp, Romatschke:2007mq, Song:2007fn} -- and
various attempts have been proposed. As part of that, the AdS/CFT correspondence
\cite{Maldacena:1997re} recently came to the forefront as it
predicts a universal bound on the ratio $\eta /{\cal S}$, with ${\cal S}$ the
entropy density, given by \cite{Policastro:2002se, Buchel:2003tz, Kovtun:2004de}
 $\eta/{\cal S} \ge 1 / 4\pi$, close to the value
obtained by fitting the relativistic heavy-ion collisions data by
hydrodynamical models and in QCD lattice simulations \cite{Meyer:2007ic}.
Moreover, it has also been applied to evaluate the jet quenching
parameter \cite{Liu:2006ug, Liu:2006nn, Liu:2006he}.

The previous ``top-down'' results are based on
the AdS/CFT gravity-gauge theory duality between strings
and supersymmetric SU(N) Yang-Mills theory in the limit
of large t'Hooft coupling. The relation of these models with  QCD
at finite temperature is suggestive, but by no means obvious.

In this paper we shall follow a ``bottom-up'' approach, instead. We ask
the double-sided question: is QCD a good {\it analog-system} of a
black-hole (BH)? or, conversely, is there a specific BH whose
thermodynamics {\it simulates} well QCD thermodynamics?


Our program starts by identifying the BH analog system
and takes as initial inputs the proposals of Ref.\cite{Castorina:2007eb}: i)
that at high energy the universal hadronic freeze-out temperature
$T_f \simeq 170$ Mev -- obtained by statistical analysis of hadronic
abundances in all collisions, $e^+e^-, pp,  p \bar p, \pi
p$, etc, including nucleus-nucleus scattering
\cite{Becattini:1995if,Becattini:1997rv,Cleymans:1992zc,Becattini:2000jw,BraunMunzinger:2003zd} --
is a Unruh temperature $T_U$
\begin{equation}\label{tasympt}
T_f|_{{\rm large} \; \sqrt{s}} = T_U = \frac{a}{2 \pi} =
\sqrt{\frac{\sigma}{2\pi}} \simeq 170 \;{\rm Mev}
\end{equation}
where $\sqrt{s} \ge 20$ Gev is the energy of the collision, $a$ is the {\it
deceleration} of quarks and antiquarks typical of the hadronic
production mechanism, and $\sigma \simeq 0.18$ Gev$^2$ is the QCD
string tension; ii) that the associated Rindler horizon can be identified with the
``color-blind'' horizon dynamically produced by the color-charge
confinement during the $q \bar{q}$ pair productions.

In this picture the hadrons produced are formed by quarks and anti-quarks,
Rindler-Unruh quanta excited out of the QCD vacuum, that are ``born in equilibrium'' (in
Hagedorn's words). This means that the hadron abundances in
the final state follow a thermal distribution not because partons  rescatter
but because of the random distribution of quarks and anti-quarks
entangled in such a vacuum. This mechanism of thermalization is encountered each time
quantum fields are near a (event) horizon, hence the
vacuum is a condensate of entangled quanta living on the two
(causally disconnected) sides (for a case simpler than QCD see,
e.g., \cite{Iorio:2004bt} and also \cite{Iorio:2001te,Iorio:2002zr}).

According to that approach the universal temperature $T_f$ found in all scattering
processes at large $\sqrt{s}$ is understood as a constant Unruh temperature. From now on we
identify $T_f$ with $T_U$. For heavy-ion collisions
$T_f$ could depend on other dynamical parameters of the produced system. For instance,
experimental data show that $T_f$ depends on the collision energy
for $\sqrt{s} \le 17$ Gev \cite{Becattini:1995if,Becattini:1997rv,Cleymans:1992zc,Becattini:2000jw,BraunMunzinger:2003zd}
and on the baryon chemical potential $\mu_b$.
These dependences are strongly correlated since the limit of large energy corresponds to zero baryon chemical
potential \cite{Becattini:2005xt}.

It is at this point the we venture into the analogy with a
gravitational system (a BH) motivated by the well-known
correspondences between acceleration, Rindler
horizon and BH horizon. Namely, we consider this hadronic Rindler spacetime as
the near-horizon approximation of some BH spacetime and pose the question: which BH?

Of course, without further input there is no unique answer to this question since many different BHs have the same near horizon approximation.
Thus what we are doing here is to look for a BH  that shares with the hadronization mechanism
certain thermodynamical properties, possibly to the extent of
enabling us to make predictions of certain behaviors, such as the
dependence of $T_f$ on the nucleus-nucleus collision energy $\sqrt{s}$
to begin with. In \cite{Castorina:2007eb} the
analogy with a Schwarzschild BH has been attempted, but the latter has unusual thermodynamical properties such as negative specific heat and does not exhibit a Hagedorn temperature. In the next section we shall identify a BH with the same near-horizon approximation, but with more
appropriate thermodynamical properties.

\section{Searching for the right black-hole}\label{se:2}

Let us first clarify that, for the hadronization processes we are
dealing with, the 2-dimensional (2D) study of the BH analog is more appropriate
than the 4-dimensional (4D) case for the following two reasons:
\begin{enumerate}
\item The dynamics of particle production is effectively 2D because it can be described in terms of the evolution
in time of the hadronic strings (string-breaking), that are one
dimensional objects.
\item The near horizon field dynamics is effectively 2D \cite{Birmingham:2001qa,Medved:2004tp,Parikh:1999mf}.
\end{enumerate}

Let us now introduce the basic ingredients of 2D dilaton gravity. It is well known that the Einstein-Hilbert action in 2D does not generate equations of motion. Dilaton gravity is the most natural generalization which leads to non-trivial dynamics\footnote{Other gravity actions, like non-linear actions in the Ricci scalar or actions which introduce torsion and/or non-metricity, can be reformulated as dilaton gravity actions, so our Ansatz is rather general. For a review cf.~e.g.~\cite{Grumiller:2002nm}.}. Its action (dropping surface terms) is
\begin{equation}\label{Action}
  I = - \frac{1}{16\pi G_2}\,\int\nts d^{\,2}x \,\sqrt{-g}\, \left[ X\,R - U(X)\,\left(\nabla X\right)^2 - \sigma_d \,V(X) \right]\,.
\end{equation}
Here $G_2$ is the 2D Newton constant which we shall set to $1/(8\pi)$ henceforth, $g$ is the metric, $R$ the associated Ricci scalar, $X$ is a scalar field (the ``dilaton'') and $\sigma_d$ is a coupling constant of dimension 1/length$^2$. The two functions $U(X)$ and $V(X)$ are unconstrained a priori and define what kind of BH solutions (if any) we obtain. The (quasi-local) thermodynamics for generic models \eqref{Action} has been extensively discussed in Ref.~\cite{Grumiller:2007ju}. We recall here some of the main results which we are going to need below.

First of all we note that there is a dimensionfull coupling constant in the action, $\sigma_d$, which controls the strength of the dilaton self-interactions. This feature is in contrast to Einstein gravity which contains no such coupling constant besides the Newton constant.
The classical solutions of the equations of motion descending from (\ref{Action})
\eq{
   X = X(r)\,, \qquad ds^2 = \xi(r) \,d\tau^2 - \frac{1}{\xi(r)}\,dr^2\,,
}{metric}
with
\eq{
  	\partial_r X =  e^{-Q(X)} \,, \qquad
	\xi(X) =  w(X) \, e^{Q(X)}\,\left( 1 - \frac{2\,M}{w(X)} \right)\,,
}{eq:defs}
are expressed in terms of two model-dependent functions,
\eq{
        Q(X) \defeq Q_0 \, + \int^{X} \bns d\tilde{X} \, U(\tilde{X})\,, \qquad
        w(X) \defeq w_0 - \sigma_d \, \int^{X} \bns d\tilde{X} \, V(\tilde{X}) \, e^{Q(\tilde{X})}\,.
}{QwDef}
Here the integrals are evaluated at $X$ and $Q_0$ and $w_0$ are two constants. However, for physical solutions a single constant of integration $M \geq 0$ is enough (cf.~e.g.~\cite{Grumiller:2002nm}) and the Ricci scalar is given by
\begin{equation}\label{eq:R}
 R = - \frac{\partial^{2}\xi}{\partial \,r^{\,2}} = -e^{-Q}\left[w''+Uw'+U'(w-2M)\right]~.
\end{equation}
For $e^Q w = 1$, $R \propto M$ and therefore the ground state solution $M=0$ is Minkowski space. We call models with this property ``Minkowskian ground state models''.


All classical solutions \eqref{metric} exhibit a Killing vector $\partial_\tau$, so we have a ``generalized Birkhoff theorem''. Therefore, each solution $X_h$ of $\xi(X_h)=0$ leads to a Killing horizon. The Hawking temperature is given by surface gravity or, equivalently, the inverse periodicity in Euclidean time,
\begin{equation}
T_{\rm Haw}=\frac{w'(X_h)}{4\pi}\,.
\label{eq:TH}
\end{equation}
For instance, when the dilaton model is the one obtained by dimensional reduction of the 4D Schwarzschild BH -- that
is we use spherical symmetry and consider the angular coordinates as spectators -- one obtains $w(X) = \sqrt{2 X / G_4}$, $X_h = 2 M^2 G_4$, where $G_4$  is the 4D Newton constant, hence $T_{\rm Haw} = w'(X_h) / 4\pi = (8 \pi G_4 M)^{-1}$, the well-known result of Hawking.


To study the thermodynamical properties (see Ref.\cite{Grumiller:2007ju}) one considers the 2D BH in a cavity with boundaries at $X=X_{\rm cav}$ in contact with a thermal reservoir at the Tolman temperature $T_{\rm cav}=T_{\rm Haw}/\sqrt{\xi(X_{\rm cav})}$ (i.e. the blue-/red-shifted temperature). In the limit $X_{\rm cav}\to\infty$, for Minkowskian ground state models, the free energy is
\begin{equation}\label{eq:FreeEnergy}
  F = M - T_{\rm Haw} \,S_{\rm BH}\,,
\end{equation}
where
\eq{
S_{\rm BH} = 2\pi\,X_h
}{eq:entropy}
is the Bekenstein-Hawking entropy \cite{Gibbons:1992rh,Nappi:1992as,Gegenberg:1994pv,Davis:2004xi,Grumiller:2006rc}, independent from the location of the cavity wall and just sensitive to local properties of the horizon, as it should be. To make contact again with well-known results we might consider once more the spherically symmetric 4D Schwarzschild BH reduced to 2D and use the previous result $X_h = 2 M^2 G_4$, which gives $S_{\rm BH} = 2\pi\,X_h = 4 \pi G_4 M^2$. Recalling that the Schwarzschild radius is $r_h = 2 G_4 M$ and that
the area of the event horizon is $A_{BH} = 4 \pi r_h^2$ we have $S_{\rm BH} = A_{BH} / 4 G_4$, the well known result of Bekenstein and Hawking.

We have now the necessary tools on the gravity side to focus on the search for our BH. Because we demand the Minkowski ground state property, the function $Q$ is determined uniquely once the function $w$ is known. Therefore, our BH is identified by constructing $w$ with the phenomenological requirements from the Rindler hadronization process. Namely we require that:
\begin{enumerate}
\item The BH mass is proportional to the energy of the collision:
\begin{equation}
    M = \gamma \sqrt{s} \,,
\end{equation}
where $\gamma$ is some numerical coefficient.

This requirement relies on the fact that, since the Hawking temperature \eqref{eq:TH} depends on the BH mass $M$, also the near horizon approximation (the Rindler description) and therefore the Unruh hadronization temperature must depend on $M$. We know that the hadronization temperature depends on energy $\sqrt{s}$, hence it is natural to identify it with $M$.

\item The coupling constant $\sigma_d$ in \eqref{Action} coincides with the string tension $\sigma$ ($\sigma = \sigma_d$).

Indeed, the string tension $\sigma$ is a fixed dimensionfull parameter, and there is only one such parameter available in \eqref{Action}, namely $\sigma_d$.

\item The Hawking temperature corresponds to the Unruh temperature,
i.e.~to the hadronization freeze-out temperature:
\begin{equation}
    T_{\rm Haw} (\sqrt{s},\sigma)= T_f(\sqrt{s},\sigma)
\end{equation}
for all values of the energy $\sqrt{s}$ and for a given value of $\sigma$.

This requirement relies on the detailed analysis of Ref.~\cite{Castorina:2007eb} already mentioned.

\item The BH partition function diverges at a given
temperature, say $T_c$ that, at $\mu_b=0$, we identify in the following way
\begin{equation}\label{tbar}
T_c ={\rm lim}_{\mu_b \to 0} T_f =
{\rm lim}_{\sqrt{s} \to \infty} T_f = \sqrt{\frac{\sigma}{2\pi}} \;,
\end{equation}
where all limits are supposed to be sufficiently smooth.

This point is motivated by the fact that massless QCD at finite temperature and $\mu_b = 0$ has a deconfining
first order phase transition.
Moreover, at zero baryon density, the
critical temperature is associated with the QCD string breaking, i.e.
with the Unruh hadronization mechanism.
Another motivation for this requirement will be given later discussing
the finite density effects.

\end{enumerate}

These points are not sufficient to identify the BH. They constrain, though, the class of allowed models severely. Point 4 in the list implies that $T_{\rm Haw}$ must be bounded from above as a function of $\sqrt{s}$. Indeed, at that value of the temperature ($T_c$) the system undergoes a phase transition and the ``hadronic Rindler horizon description'' is no longer applicable, hence our BH analog description also must break down there. This requirement excludes most of the well-known BHs, such as Schwarzschild or Reissner-Nordstr\"om in any dimension which have no such a critical temperature. Furthermore, from the phenomenological analysis of the nucleus-nucleus scattering \cite{Becattini:2005xt} the behavior of $T_f$ at large but finite $\sqrt{s}$ turns out to be
\eq{
T_{\rm Haw} = T_f \simeq  T_c\left(1 - \frac{\sqrt{s_0}}{\sqrt{s}} + {\cal O}(1/s)\right)\,.
}{eq:fifth}
This is consistent with the fourth requirement, but slightly stronger because it contains also information about the next to leading order term in a large $s$ expansion.

Let us now consider first the leading order term $T_c$. Noticeably, this establishes a \textit{unique} asymptotic BH model: since $T_{\rm Haw}$ to leading order must be given by the constant $T_c$ we can deduce from \eqref{eq:TH} that the function $w$ must be linear in $X$ in the limit of large $M$. The unique BH model which does the job is known as ``Witten BH'' \cite{Witten:1991yr},
\eq{
w(X) = \sqrt{8\pi\sigma_d}\, X\,,
}{eq:WBH}
and arises as an approximate solution in 2D string theory to lowest order in $\alpha^\prime$. Its Hawking temperature is then $T_{\rm Haw} = \sqrt{\sigma_d / 2 \pi}$ and, by the second and third phenomenological requirements we get
\eq{
T_{\rm Haw} = T_f = \sqrt{\frac{\sigma}{2 \pi}} \;.
}{eq:TWBH} 
Furthermore, the partition function of the Witten BH diverges\footnote{As discussed in \cite{Davis:2004xi} we consider the Witten BH in a cavity whose wall is
located at some fixed value of the dilaton $X=X_{\rm cav}$. The cavity is in
contact with a thermal reservoir at $T=T_{\rm cav}$. Allowing for all paths
where the metric is continuous (but not necessarily differentiable)
gives a Euclidean partition function $Z$ which is an infinite sum over
instantons. Most of them exhibit a conical defect \cite{GMinpreparation}. 
For the Witten BH
the resulting integral can be exactly solved, giving $Z \propto X_{\rm cav}$
for very large $X_{\rm cav}$. Eventually we move the cavity wall to infinity --
because that is where the asymptotic observer sits, measuring the
Hawking temperature -- and this means that $Z \to + \infty$.
Physically, the reason for this divergence is the singular specific heat
of the Witten BH, i.e., the divergence of fluctuations. Another way to put
it is to observe that the Witten BH is marginally unstable against decay
into conical defects \cite{GMinpreparation}.}, hence in
particular it
is divergent at $T_f$. This we regard as an instance of the
fulfillment of the fourth requirement.
All this leads us to recognize the Witten BH as the unique BH reproducing the behavior of the freeze-out temperature for all the scattering processes at high energy considered in \cite{Castorina:2007eb} except heavy-ion collisions ($e^+ e^-$, $p p$, $p \bar{p}$, $\pi p$, etc.). That is to say that the Witten BH is the BH we wanted in all the high energy scattering processes when the freeze-out temperature is constant and equal to the critical temperature.

For heavy-ion collisions we are therefore looking for a deformation of the Witten BH that, at finite values of $s$, is consistent with \eqref{eq:fifth}.Since the Witten BH emerged as the unique approximation to lowest order in $\alpha^\prime$, the only natural candidate is the exact solution in 2D string theory to all orders in $\alpha^\prime$, which is known as the ``exact string BH'' \cite{Dijkgraaf:1992ba}. Its target space action was constructed in \cite{Grumiller:2005sq}. Like the Witten BH it is a Minkowskian ground state model given by
\eq{
w(X) = \sqrt{8\pi\sigma_d} \,\left(\sqrt{\rho^2+1}+1\right)\,,\qquad e^{Q(X)} w(X) = 1\,,
}{eq:EBH1}
where the canonical dilaton $X$ is related to a new field $\rho$ by
\eq{
X=\rho+\arcsinh{\rho} ~.
}{eq:EBH2}
Obviously, for $X\to\infty$ \eqref{eq:EBH1} with \eqref{eq:EBH2} asymptotes to \eqref{eq:WBH}.

Its Hawking temperature is given by \cite{Grumiller:2005sq}
\eq{
T_{\rm Haw} = \sqrt{\frac{\sigma_d}{2\pi}}\,\sqrt{1-\frac{2 \sqrt{2 \pi \sigma_d}}{M}}\,.
}{eq:THE}
where $M$ is the ADM mass of the BH\footnote{The mass $M$ is also related to the level $k$ of the current algebra underlying the CFT description of the exact string BH in terms of an $SL(2,\BR)/U(1)$ gauged WZW model (for a review cf.~e.g.~\cite{Becker:1994vd}). That $k$ can be seen as a running parameter allowed by the CFT is discussed in \cite{Kazakov:2001pj}.}.


Thus the exact string BH is a 2D BH fulfilling all the phenomenological requirements: 1.~The first condition is satisfied by identifying $\gamma \sqrt{s} = M$; 2.~The second requirement is simply $\sigma = \sigma_d$; 3.~The third postulate will allow us to make predictions about $T_f(\sqrt{s},\sigma)$, which we shall discuss in Section \ref{se:4}; 4.~The fourth postulate is met, because \eqref{eq:THE} obviously is bounded from above by \eqref{tbar}.

Having demonstrated that the exact string BH is phenomenologically viable, we address now the issue of uniqueness. We have shown above that asymptotically (for large $s$) the Witten BH emerges as the {\em unique} BH model consistent with all requirements. While there is a whole family of models that asymptotes to the Witten BH, we have also noted that from a CFT point of view there is a {\em unique} BH model that deforms the Witten BH for finite values of $s$, namely the exact string BH. In that sense our results are unique.

\section{Matching the phenomenological results for $T_f (\sqrt{s})$}\label{se:4}

\begin{figure}
{{\epsfig{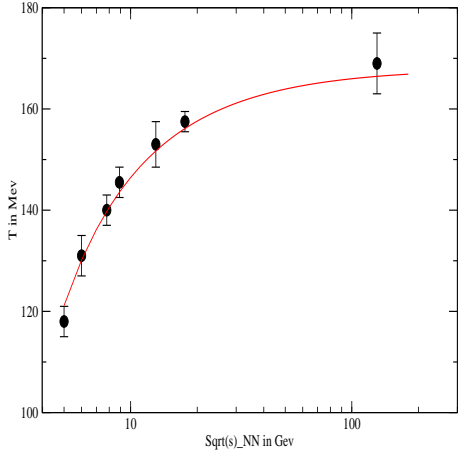}} \caption{Freeze-out temperature versus $\sqrt{s}_{NN}$
from Ref.~\cite{Becattini:2005xt} compared with Eq.~(\ref{tfesbh}) for
$\sqrt{s_0}= 2.4$ Gev and $T_c = 169$ MeV.}}
\end{figure}

From the above discussion Eq.~(\ref{eq:THE}) leads to the following prediction for the energy dependence of the
freeze-out temperature in heavy ion collisions
\begin{equation}\label{tfesbh}
T_f(\sqrt{s}) =  \sqrt{\frac{\sigma}{2\pi}} \sqrt{1 - \sqrt{\frac{s_0}{s}}} \;,
\end{equation}
where $\sqrt{s_0} = 2 \sqrt{2 \pi \sigma} / \gamma $ is a free parameter.
In Fig.~1 we compare Eq.~(\ref{tfesbh}) with the phenomenological results of Ref.~\cite{Becattini:2005xt}
for the dependence of the freeze-out temperature on the collision energy, for different nuclei, for $\sqrt{s_0}=2.4$ Gev. At large energy
the universal value $T_f \simeq 170$ Mev  is obtained.

As previously discussed, the $\sqrt{s}$ dependence of $T_f$ is strongly correlated
with its dependence on the baryon chemical potential $\mu_b$ \cite{Becattini:2005xt}. We stress that the $\mu_b$-dependence of $T_f$ is different from the
the $\mu_b$-dependence of $T_c$, where the corresponding
BH partition function diverges \cite{Magas:2003wi}.
Indeed, at $\mu_b=0$, the deconfinement temperature is related, in the Hagedorn model \cite{Hagedorn:1965st} or
in the dual resonance model \cite{Fubini:1969qb, Bardakci:1970yd}, to the resonance formation
and decay and therefore string-formation and  -breaking is the
relevant dynamical mechanism. This means that, at $\mu_b=0$, it is reasonable to
consider the freeze-out temperature essentially equal to the temperature at the point of
deconfinement. This is another  motivation for our assumption 4 in the
previous list.
At finite $\mu_b$, the interaction does not lead to the formation of resonances
but the screening effects and Fermi statistic (at large
$\mu_b$) play the most important role. Hence there is no a priori
reason for $T_f \simeq T_c$.
Accordingly, in the language of the 2D BH thermodynamics the
$\mu_b$-dependence of $T_c$ should be studied
by introducing in the dynamics a new conserved $U(1)$ charge,
corresponding to the baryon number, and considering the critical
line in the $T-\mu_b$ plane.

\begin{figure}
{{\epsfig{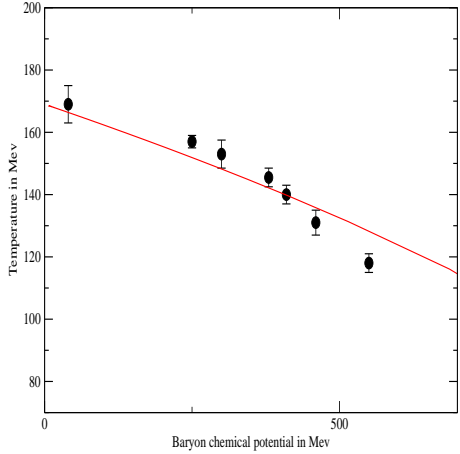}} \caption{Freeze-out temperature versus  baryon chemical
potential  compared with Eq.~(\ref{tfmub}) for $\mu_b^0 = 1.2$ GeV.}}
\end{figure}

We now come back to the $\mu_b$-dependence of $T_f$. At a purely phenomenological level it can be described by the empirical relation
\begin{equation}\label{density}
\mu_b \simeq \frac{c}{\sqrt{s}} \;,
\end{equation}
where $c$ is a constant. The approximate relation \eqref{density} comes from the statistical analysis of the species abundances in heavy
ions collisions \cite{Becattini:2005xt}. Thus, by inserting (\ref{density}) into (\ref{tfesbh}) we obtain
\begin{equation}\label{tfmub}
T_f(\mu_b) \simeq  \sqrt{\frac{\sigma}{2\pi}} \sqrt{1 - \frac{\mu_b}{\mu_b^0}} \;,
\end{equation}
where $\mu_b^0$ is a free parameter. In Fig.~2 we compare the prediction of Eq.~(\ref{tfmub}), for
$\mu_b^0 = 1.2$GeV, with the phenomenological analysis. Our curve merely gives a rough estimate of the $\mu_b$-dependence of $T_f$
and could be regarded as a theoretical prediction that is only indirectly based on the BH analogy, i.e. via Eq.~(\ref{density}). Keeping
these limitations in mind, we nonetheless notice that Eq.~(\ref{tfmub}), if taken at face value, predicts a linear
screening of the string tension due to finite density effects in heavy ions collisions
\begin{equation} \label{screening}
\sigma (\mu_b) \simeq \sigma (1 - \mu_b / \mu_b^0 ) \;.
\end{equation}
This behavior of $\sigma$, being $\mu_b^0=1.2$GeV, gives as critical quark chemical potential $\mu^0_q \simeq 400$ MeV.

\section{Conclusions}\label{se:3}

In this work we have further investigated the recent proposal that for all high energy hadron productions the universal hadronic freeze-out
temperature, $T_f \simeq 170$ Mev, can be understood as a Unruh temperature. 
Here we identified the exact string BH (for the heavy-ion collisions) and its limiting case, the Witten BH (for all the other processes), as the unique BHs whose thermodynamical properties well simulate some thermodynamical properties of hadronization. In particular, exploiting the behavior of Hawking temperature for the exact string BH we provided an analytical expression for the energy dependence of the freeze-out temperature in nucleus-nucleus scattering which gives a very good fit of the experimental data obtained via the statistical
hadronization model. We also proposed a linear screening of the hadronic string tension as a function of the baryon chemical potential based on an empirical relation.

In view of taking this work as a first step of a bottom-up program of finding a BH
whose thermodynamics could simulate finite temperature QCD, it is perhaps suggestive to recall
here some of the thermodynamical properties of the exact string BH:
The partition function diverges at $T_c$; The specific heat is positive; The third law of thermodynamics holds,
i.e. the specific heat vanishes linearly with temperature as the latter approaches zero.

Finally, let us merely report here the following coincidence. Besides the exact string BH that we advocate here
there is another BH that has been applied to QCD, namely the well-known BH in $AdS_5$. We can look at the near singularity behavior of the $AdS_5$ BH and compare it with the near singularity behavior of the T-dual of the exact string BH. It turns out \cite{Grumiller:2005sq}, that they have the same behavior.





\section*{Acknowledgments}

We thank Ji\v{r}i Ho\v{s}ek, Hong Liu and Anton Rebhan for discussion. 

P.C.~and D.G.~acknowledge the kind hospitality of the Institute for Particle and Nuclear Physics of Charles University of Prague, and A.I.~acknowledges the kind hospitality of the Department of Physics and Astronomy of Catania University and of the Center for Theoretical Physics of MIT.

P.C.~has been supported in part by the INFN-MIT ``Bruno Rossi'' program. D.G.~is supported in part by funds provided by the U.S.~Department of
Energy (DoE) under the cooperative research agreement DEFG02-05ER41360 and by the project MC-OIF 021421 of the European Commission under FP6. A.I.~has been supported in part by the Department of Physics ``Caianiello'' and INFN, Salerno University and by the project MC-OIF 021421 of the European Commission under the Sixth EU Framework Programme for Research and Technological Development (FP6).


\providecommand{\href}[2]{#2}\begingroup\raggedright\endgroup

\end{document}